# An Infinite Number of Closed FLRW Universes for Any Value of the Spatial Curvature[1]

Helio V. Fagundes[2]
Instituto de Física Teórica, Universidade Estadual Paulista
São Paulo, SP, CEP 01140-070, Brazil

**Abstract**

The Friedman-Lemaître-Robertson-Walker (FLRW) cosmological models are based on the assumptions of large-scale homogeneity and isotropy of the distribution of matter and energy. They are usually taken to have spatial sections that are simply connected; they have finite volume in the positive curvature case, and infinite volume in the null and negative curvature ones. I want to call the attention to the existence of an infinite number of models, which are based on these same metrics, but have compact, finite volume, multiply connected spatial sections. Some observational implications are briefly mentioned.

## I. Friedman-Lemaître-Robertson-Walker Models

As is well known, the Friedman-Lemaître-Robertson-Walker (FLRW) cosmological models have the general metric - see, for example, [1] -

$$ds^2 = dt^2 - a^2(t)\left(\frac{dr^2}{1-kr^2} + r^2 d\theta^2 + r^2 \sin^2\theta\, d\varphi^2\right),$$

with spatial sections $S^3(t)$ for $k = 1$, $E^3(t)$ for $k = 0$, and $H^3(t)$ for $k = -1$. The spherical sections have the finite volume $2\pi^2 a^3(t)$, while the Euclidean and hyperbolic ones have infinite volumes. When I first studied this subject, I was uncomfortable with this situation.

Then I learned, from Efimov's *Higher Geometry*[3] [2], that the $H^2$ plane could be tessellated into a mosaic of regular, $4g$-sided, hyperbolic polygons, where $g \geq 2$; the tessellation (or 'honeycomb') is generated by a symmetry group $\Gamma_g$ such that $H^2/\Gamma_g$ is a *closed* – meaning *compact and without boundary* – surface of genus *g*. In my first paper on this matter [4], I took advantage of this fact to construct a closed cosmological model with sections $\left(H^2/\Gamma_g\right)\otimes S^1$, so I had rediscovered one of Kantowski-Sachs metrics [5]. But these sections could approximate the $H^3(t)$ metrics only in a small range of *t*.

---

[3] As a graduate student at the California Institute of Technology (CALTECH), I was free to roam through its library's bookshelves. There I found Efimov's treatise quoted above and A. P. Norden's elegant textbook on Lobachevskian geometry [3]. I eagerly studied the latter, despite my poor German.

## II. J. Wolf's Book and L. Best's Paper

In Wolf's *Spaces of Constant Curvature* [6], I found infinitely many compact, orientable 3-manifolds with *k* = 0 and *k* = 1. There are six classes of such Euclidean spaces, and a countably infinite number of spherical ones. Except for spherical space itself, these compact spaces have nontrivial topology, their fundamental group being the tessellating group. E. g., for the flat torus $\pi_1 = \mathbf{Z}^3$, for projective space $\pi_1 = \mathbf{Z}_2$. But Wolf's space forms include the ones with hyperbolic metric, because he restricted his book to *homogeneous*, compact manifolds and there is a theorem (cf. [7], for example) that asserts there are no such manifolds of negative curvature.

Then I read a paper by L. A. Best [8], where a compact, hyperbolic 3-space was constructed, what led to my publishing of a two-page article [9], with that manifold as spatial section[4]. There it was shown that, if homogeneity of the *space of repeated images* was substituted for homogeneity of the sources' distribution, then the model would not contradict, on this matter, the data of extra-galactic astronomy.

Best's paper also presented a number of closed hyperbolic 3-manifolds, based on a regular icosahedron and a regular dodecahedron as fundamental tiles.

## III. Weeks's *SnapPea* Software

My final source for hyperbolic, closed 3-spaces came from American geometer J. R. Weeks, who had done his Ph. D. work at Princeton University, under Fields medalist William P. Thurston, known by mathematicians for his work on the topology and geometry of 3-manifolds.

---

[4] A theoretical background for the use of multiply connected spaces in FLRW models has been given by Ellis [10].

Weeks's thesis was the development of a computer program, which he called *SnapPea* [11], for the construction and study of such spaces, based on their connection with the theory of knots. This executable program contained a catalog of about 11,000 entries, ordered by increasing volume (with the sectional curvature normalized to $K = -1$), starting with one of volume = $0.94…$; and it provided mathematical data on each space, such as the Lorentz representation in Minkowski coordinates, for the generators of its tessellation.

Jeff sent me a copy of *SnapPea*, which was originally written only for Macintosh computers. I got a US$2,500 grant from Fundação de Amparo à Pesquisa do Estado de São Paulo, and Government permission, to import one such machine for Instituto de Física Teórica of Universidade Estadual Paulista (São Paulo), so my students and I could have access to the program.

## IV. An isosahedron-based cosmological model

Let me finish by mentioning one of my favorite works, the article *Quasar-Galaxy Associations with Discordant Redshifts as a Topological Effect. II. A Closed Hyperbolic Model* [12].

It had one of Best's icosahedron-based manifolds as spatial section. The title refers to an application that tried to solve a then famous controversy on the nature and distance of quasi-stellar sources or quasars. My paper showed the possibility of the image of a galaxy being aligned with much more distant, and hence earlier, images of the same galaxy − when it could have been a quasar. Then H. Arp's quasar-galaxy associations [13] might be produced by this topological process, instead of his same-distance interpretations.

This model was constructed before the work of the just announced physics 2011 Nobel Prize winners. It assumed $\Lambda = 0$, $\Omega = 0.30$, the then favorite values among astrophysics, and a number of source-second image alignments was obtained. These results were not good enough for comparison with astronomical observations, but the methods used for dealing with hyperbolic geometry might still be useful.

As a final remark, I refer to applications of the compactness of cosmological space in my 2001 paper [14].